# Development of depleted monolithic pixel sensors in 150 nm CMOS technology for the ATLAS Inner Tracker upgrade

**P. Rymaszewski**[*a], **M. Barbero**[b], **S. Bhat**[b], **P. Breugnon**[b], **I. Caicedo**[a], **Z. Chen**[b], **Y. Degerli**[c], **S. Godiot**[b], **F. Guilloux**[c], **C. Guyot**[c], **T. Hemperek**[a], **T. Hirono**[a], **F. Hügging**[a], **H. Krüger**[a], **M. Lachkar**[c], **P. Pangaud**[b], **A. Rozanov**[b], **P. Schwemling**[c], **M. Vandenbroucke**[c], **T.Wang**[a], **N. Wermes**[a]

[a] *University of Bonn,*
 *Nussallee 12, Bonn, Germany*
[b] *Centre de physique des particules de Marseille,*
 *163 Avenue de Luminy, Marseille, France*
[c] *IRFU, CEA-Saclay,*
 *Gif-sur-Yvette Cedex, 91191 France*

*E-mail:* rymaszewski@physik.uni-bonn.de

This work presents a depleted monolithic active pixel sensor (DMAPS) prototype manufactured in the LFoundry 150 nm CMOS process. The described device, named LF-Monopix, was designed as a proof of concept of a fully monolithic sensor capable of operating in the environment of outer layers of the ATLAS Inner Tracker upgrade for the High Luminosity Large Hadron Collider (HL-LHC). Implementing such a device in the detector module will result in a lower production cost and lower material budget compared to the presently used hybrid designs. In this paper the chip architecture will be described followed by the simulation and measurement results.

---

[*]Corresponding author.

# 1. Introduction

During the Phase-II upgrade of the ATLAS detector the whole inner tracker will be replaced with silicon detectors, where the first five layers will be made from pixel devices [1]. Such an action is necessary due to envisioned increase of luminosity of the LHC, which in turn will significantly increase the radiation levels in the detector and pose new challenges to readout electronics due to the increased particle rate.

Presently the baseline design for pixel layers is the hybrid approach, where the silicon sensor is bump bonded to a dedicated readout chip. However, within the ATLAS Inner Tracker (ITk) collaboration, a programme called "CMOS demonstrator" was started with a goal of evaluating the use of commercial CMOS technologies to build silicon pixel detectors, especially for outer pixel layers where the devices need to withstand 80 Mrad TID and the fluence of $1.5 \cdot 10^{15} \left[\frac{n_{eq}}{cm^2}\right]$. One of the evaluated approaches is a monolithic design, where the sensor and full readout electronics are integrated into one chip, resulting in simpler assembly and lower cost of detector modules.

Though there are already high energy physics experiments using monolithic active pixels [2][3], the presently available devices rely on diffusion for charge collection, which is slow (order of a few $\mu$s) and therefore unsuitable for ATLAS requirements (25 ns time resolution). To overcome this limitation, the "CMOS demonstrator" community is trying to exploit CMOS technology add-ons (high resistive substrate, multiple nested wells) and apply high bias voltage to deplete the sensing volume to collect the charge with drift, which is much faster (order of ns). Devices following this approach are called depleted monolithic active pixel sensors (DMAPS) and the presented prototype, named LF-Monopix, is a proof of concept of a DMAPS suitable for the outer pixel layers of a Phase-II ATLAS upgrade.

# 2. Design description

The device presented in this paper has been manufactured using LFoundry 150 nm CMOS technology on a high resistivity ($> 2k\Omega$·cm) P-type wafer. The charge collection electrode is created by the very deep N-well, which also encapsulates all the pixel electronics (a simplified pixel cross section is presented in Figure 1). Consequently, LF-Monopix has a large fill factor (ratio of the collecting electrode to the area of the pixel), resulting in high sensor capacitance ($\approx$ 400 fF) and potentially more radiation hardness than the small fill factor devices. Use of both NMOS and PMOS transistors in the pixel is enabled by the presence of a deep P-well, which shields the transistor's N-well from the charge-collecting very deep N-well. Many aspects of the design (e.g., charge sensitive amplifiers (CSA), one of two comparator types, configuration registers and DACs) are inherited from previous prototypes made in this process [4][5]. The novel part in this chip is the addition of a fully integrated fast readout, which uses a column drain architecture very similar to the one implemented in FEI3 chip (currently used in ATLAS pixel detector [6]).

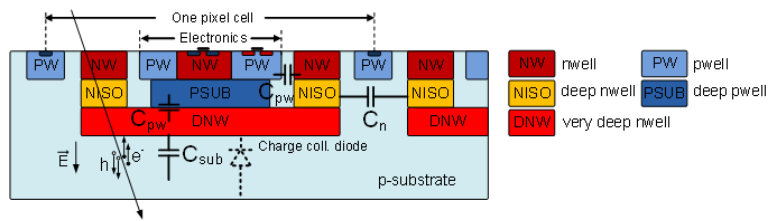

**Figure 1:** Generalised cross section of the pixel used in LF-Monopix. In addition, the main parasitic capacitances connected with the sensing node are indicated.

The chip size is 1 cm $\times$ 1 cm, where most of the area is occupied by the pixel matrix composed of 36 columns and 129 rows. The size of the implemented pixel is 250 $\mu$m $\times$ 50 $\mu$m. In every pixel, the charge signal is first amplified by a CSA and then compared to an adjustable global threshold by a comparator (a 4-bit in-pixel DAC can be used to correct for threshold dispersion). Two 8-bit time stamps corresponding to the leading and trailing edge of the comparator output, respectively, are stored in SRAM cells, thus the time of arrival and time over threshold (ToT) are obtained. A hit in a pixel initiates token signal propagation to a readout controller, which starts the priority scan, such that hit pixels are successively read out. The data is serialised and sent off-chip at a rate of 160 Mbps by a LVDS driver. For design simplicity, the readout controller is implemented off-chip in a FPGA. The chip block diagram with a pixel schematic is presented in Figure 2.

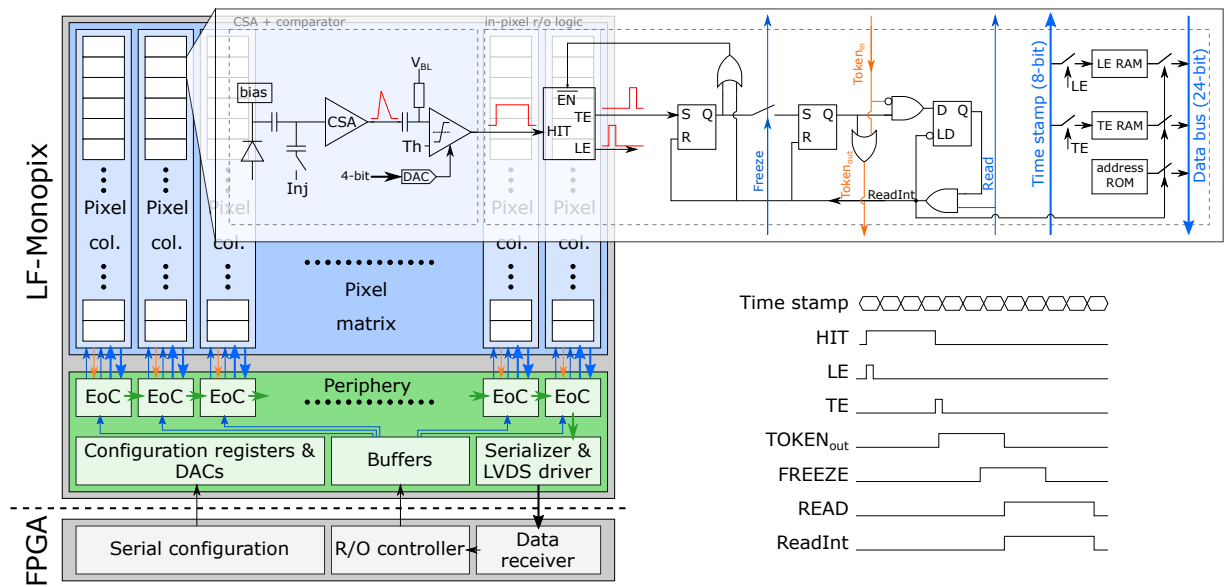

**Figure 2:** Chip architecture with a schematic of in-pixel electronics (signal lines in blue are common column-wide). EoC stands for end-of-column circuit, which includes Grey coded time stamp generation from an externally provided clock signal (40 MHz by default), sense amplifiers for reading memories, and digital logic for priority arbitration and data propagation. The presented waveforms show the readout procedure following the column drain readout approach.

The pixel matrix is not uniform and includes nine pixel flavours (four columns per flavour). The pixel types are composed of different combinations of implemented CSA architectures (two variants: NMOS transistor input and CMOS input), comparator architectures (two types: two-stage open-loop structure and a self-biased differential amplifier with a CMOS output stage), a type of logic gate propagating token signal (CMOS or current steering gate), and placement of in-pixel readout logic (either directly inside the pixel or at the periphery with one-to-one connection between the pixel and logic cell). Power consumption for all flavours is $\approx 36~\mu$W per pixel with front ends optimised for peaking time < 25 ns and ENC $\approx 200$ e$^-$. A more detailed description of pixel flavours (including techniques used for noise minimisation, e.g., current steering logic and source follower SRAM readout) togehter with more in-depth simulation results can be found in [7].

## 3. Measurement results

The described prototype was received back from fabrication in March 2017, and measurement efforts are still on-going. The breakdown voltage was measured at room temperature for a few chips, and for all of them breakdown occurs at approx. 280 V. This result is an improvement over previous prototypes and will most likely allow for efficient charge collection even after irradiation (previous prototypes with a very similar sensor design showed a 50 $\mu$m depletion depth and high efficiency at 100 V bias after NIEL fluence of $10^{15} \left[\frac{n_{eq}}{cm^2}\right]$ [8]).

The internal injection circuit was used to calibrate the ToT value as a function of the collected charge. After obtaining this information, LF-Monopix was exposed to a $^{241}$Am source (59.5 keV) and an X-ray flu-orescence from $^{65}$Tb ($K_\alpha$: 44.5 keV, $K_\beta$: 50.4 keV), which allowed measuring the real injection capacitance value (≈2.7 fF) after peak fitting (presented in Figure 3). This value is relevant, as it will be used for future voltage-to-charge calibrations and threshold determination.

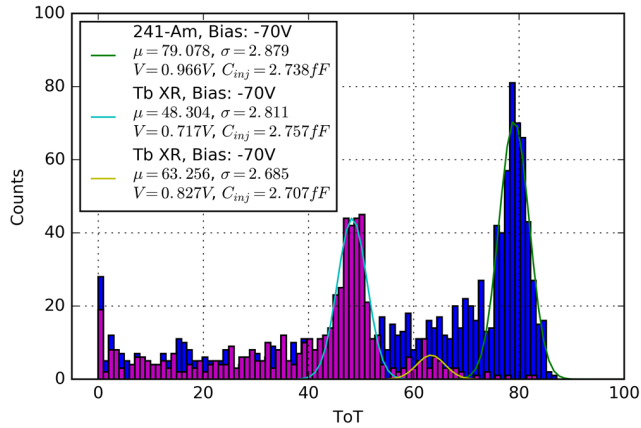

**Figure 3:** Single-pixel spectra of two sources with fit curves.

Presently, not all pixel flavours have been characterised. However, the threshold distribution and ENC for two pixel variants are presented in Figure 4. Both use CMOS input CSA with readout logic inside the pixel: variant A includes a two-stage open-loop gain comparator used in previous prototypes, and variant B uses a new comparator design - a self-biased differential amplifier. In Figure 4(a), the new comparator exhibits a larger dispersion, which was traced back to smaller input transistors than those in the old comparator. A larger dispersion with a tuning algorithm that is not fully optimised leads to different tuned threshold values. Further improvements of the tuning algorithm will most likely result in an even lower threshold than 2.5 ke$^-$. The noise distribution shown in Figure 4(b) is very similar for both pixel flavours, as expected, and the mean value ≈ 200 e$^-$ is a good match to the simulation results.

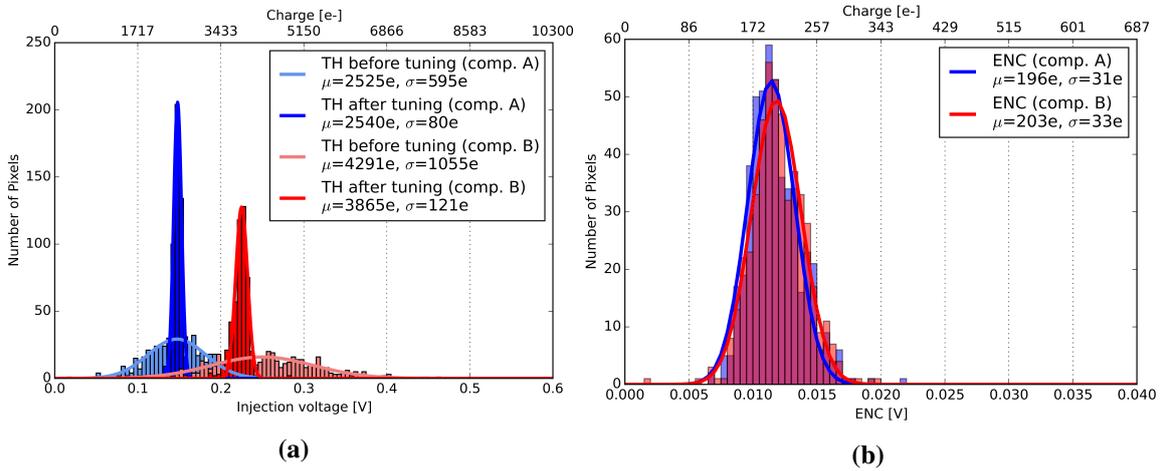

**Figure 4:** Measured (a) threshold distribution and (b) noise distribution for two pixel flavours.

First results from a sample neutron-irradiated to the fluence of $10^{15} \left[\frac{n_{eq}}{cm^2}\right]$ are positive, but the full characterisation of the device is not finished. The communication with the chip is possible, and the monolithic readout is working correctly. The leakage current increased as expected, and the breakdown voltage remains above 200 V. Further investigation will follow in the coming weeks, with test beam campaigns to determine the efficiency of this prototype.

## 4. Summary

This paper presents a DMAPS prototype with integrated fast readout logic. The device was manufactured using the LFoundry 150 nm CMOS technology on a high resistivity wafer. Measurement results show that the chip is fully functional and behaves as predicted by simulations. Breakdown voltage was measured at 280 V, and the threshold of 2.5 ke$^-$ with dispersion of 100 e$^-$ was achieved using a monolithic readout with 25 ns time stamp resolution. Measurements of efficiency for both non-irradiated and irradiated to a NIEL fluence of $10^{15} \left[\frac{n_{eq}}{cm^2}\right]$ devices are ongoing.

## Acknowledgments

This project has received funding from the European Union's Horizon 2020 Research and Innovation programme under Grant Agreements no. 654168 and 675587-STREAM.